\documentclass[aps,twocolumn,showpacs,preprintnumbers,amsmath,amssymb]{revtex4}
\usepackage{dcolumn}
\usepackage[dvips]{epsfig}

\begin{document}
\title{\bf  Black hole entropy arising from massless scalar field with Lorentz violation induced
by the coupling to Einstein tensor}

\author{Songbai Chen\footnote{csb3752@hunnu.edu.cn},  Jiliang Jing\footnote{jljing@hunnu.edu.cn}, Hao Liao
}

\affiliation{Institute of Physics and Department of Physics, Hunan
Normal University,  Changsha, Hunan 410081, People's Republic of
China \\ Key Laboratory of Low Dimensional Quantum Structures \\
and Quantum Control of Ministry of Education, Hunan Normal
University, Changsha, Hunan 410081, People's Republic of China\\
Synergetic Innovation Center for Quantum Effects and Applications,
Hunan Normal University, Changsha, Hunan 410081, People's Republic
of China\\State Key Laboratory of Theoretical Physics, Institute of Theoretical Physics, Chinese Academy of Sciences,
Beijing 100190, China
\\Kavli Institute for Theoretical Physics China, CAS, Beijing 100190,China}

\begin{abstract}
\begin{center}
{\bf Abstract}
\end{center}

We have investigated quantum entropy of a static black hole arising
from the massless scalar field with Lorentz violation induced by the
coupling to Einstein tensor. Our results show that the coupled
massless scalar field contributes to the classical
Bekenstein-Hawking term in the black hole entropy. The corrected
classical Bekenstein-Hawking entropy is not one quarter of the event
horizon area of the original background black hole, but of a
corresponding effective metric related to the coupling. It means
that the classical Bekenstein-Hawking entropy depends not only on
the black hole parameter, but also on the coupling which reduces
Lorentz violation.

\end{abstract}

\pacs{ 04.70.Dy, 95.30.Sf, 97.60.Lf } \maketitle
\newpage

Black hole entropy, which was first introduced by Bekenstein and
widely accepted after Hawking's discovery of the black hole thermal
radiation \cite{Beken1,Hawk1,Beken2},  plays an important role in
the modern physics because it is connected with gravity, quantum
theory, thermodynamics and statistics. However, the exact physical
origin of the black hole entropy and the degrees of freedom counted
for the entropy are still unclear. The resolutions of such
fundamental problems are extremely important for black hole physics.
To obtain a microscopic statistical interpretation to the black hole
entropy, 't Hooft \cite{Hooft} proposed the ``brick wall" model in
which the black hole entropy is identified with the
statistical-mechanical entropy which arises from a thermal bath of
quantum fields propagating outside the event horizon. In this model,
two extra ``brick wall" cut-off factors $h$ and $L$ are introduced to
eliminate the ultraviolet divergence near the event horizon and the
infrared divergences at the spatial infinite, respectively. It is equivalent to imposing Dirichlet boundary condition on the boundary. Combining with the
lowest order (in terms of $\hbar$) Wentzel-Kramers-Brillouin (WKB)
approximation, one can find that the quantum corrected entropy of a
black hole is approximated as
\begin{eqnarray}
S=\frac{A_H}{48\pi
\epsilon^2}+\mathcal{G}(A_H)\log\frac{\Lambda}{\epsilon},
\end{eqnarray}
where quantities $\epsilon$ and $\Lambda$ are related to the cut-off
factors $h$ and $L$, respectively. The factor $\mathcal{G}(A_H)$ is
a function of the event horizon area $A_H$. It means that quantum
entropy of a black hole contains a usual Bekenstein-Hawking term and
a logarithmic correction term. The former can be expressed as one
quarter of the event horizon area by the renormalization of the
Newton gravity constant, which is independent of the quantum fields
propagating in the spacetime. The latter describes the contribution
of quantum fields to the black hole entropy, which depends heavily
on the behaviors of the quantum fields and the properties of the
background spacetime. Both of them have the factor of the event
horizon area, which means that black hole entropy possesses the
geometric character. Thus, there's no doubt that the ``brick wall"
model reveals some remarkable features of quantum entropy of black
holes. Meanwhile, it is worth to  note that as a semiclassical
approach, the ``brick wall" model has some drawback. For example, it
has to impose two cut-off factors to eliminate the divergence near
the event horizon and at the spatial infinite, which results in  that
the ``brick wall" entropy depends on the cut-off scale. Moreover,
due to adopting the lowest order WKB approximation,  densities of the
states and canonical entropy of the matter fields are usually
evaluated with no higher precision. However, it is obvious that one
can obtain the area law of Bekenstein-Hawking entropy
through the ``brick wall" model. Thus, this approximate model has
been extensively used to compute the leading order terms in the
black hole entropy for different theories of gravity.

In cosmology, it is well known that the extra interaction terms are
introduced in the Lagrange action to explain the accelerating
expansion of the current Universe. It is widely believed that the accelerating expansion of the Universe confirmed by the current observations
\cite{ae1,ae2,ae3,ae4} will exert a profound influence upon the
modern cosmology and theory of gravity. One of the most interesting
interactions is the coupling between scalar field and the Einstein
tensor, which is described by the term
$G^{\mu\nu}\partial_{\mu}\psi\partial_{\nu}\psi$ \cite{Sushkov}. The
attractive features of such theory are that the derivative coupling
term can provide not only a mechanism to explain the accelerating
expansion of the current Universe, but also a mechanism to solve
naturally the problem of a graceful exit from inflation without any
fine-tuned potential in the early Universe
\cite{Sushkov,Sarida,Gao,CG1,James}. Moreover, one can find that in this
theory the equation of motion  of scalar field is still a
second-order differential equation, which means that it is a ``good"
dynamical theory \cite{Sushkov}.

From the point of view of physics, a good theoretical model in
cosmology should be examined by black hole physics since black hole
is another fascinating aspect in the modern theories of gravity.
Thus, it is very necessary to study such coupling theory in black
hole physics. We \cite{sb2010} studied the dynamical evolution for a
scalar field coupling to Einstein's tensor in the background of
a Reissner-Nordstr\"{o}m black hole and found a distinct
behavior of the coupled scalar field that the growing modes appear
as the coupling constant is larger than a certain threshold value.
Moreover, Minamitsuji \cite{Mas} investigated such coupling
theoretical model in Braneworlds
 and the scalar-tensor theory, and found that a new phase transition from a Reissner-Nordstr\"{o}m
black hole to a hairy black hole takes place in asymptotically flat
spacetime because the Abelian $U(1)$ gauge symmetry is broken in the
vicinity of the black hole horizon when the coupling constant is
large enough \cite{BHC2,Theod,BHC1}. These results have attracted more
attention on the investigations of the full properties for such
special coupling model in the black hole physics.

Since the coupling term modifies the behaviors of scalar fields
propagating in the spacetime, it is expected that the coupling will affect
black hole entropy. However, unlike in the usual case, the coupling
between massless scalar field and Einstein tensor changes the
dispersion relations and the maximum propagation velocity of the
field, which yields the coupled
massless scalar field does not satisfy Lorentz invariance again.
It means that Lorentz invariance is broken as the scalar field couples to Einstein tensor.
Recently, Lorentz violation has been investigated extensively in different theoretical
models because its existence will have a great influence on the fundamental
physics and yield that many subjects are need to be reconsidered (see
\cite{LI0} for a review). Therefore, it is  quite necessary to study
quantum entropy of black hole arising from the massless scalar field
with Lorentz violation induced by the coupling to Einstein tensor.
However, the appearance of Lorentz violation make the calculation of
black hole entropy more complicated. Firstly, for the coupled massless scalar field, the position of the event
horizon of a black hole is not
overlapped again with that for the light because
its propagation velocity in the background
spacetime is not equal to the speed of light. Moreover, the usual method
for calculating quantum entropy of black hole arising material field
is invalid because the statistical mechanics must be corrected
in the case with Lorentz violation \cite{LI}.
Fortunately, such Lorentz violation induced by the coupling to
Einstein tensor can be regarded formally as Lorentz invariance in
another effective metric, where the coupled massless scalar field
in the original background spacetime can be looked as a free scalar
field propagating with the speed of
 ``light". Thus, in the
 effective metric, Lorentz invariance is guaranteed formally for the scalar
 field and calculating quantum entropy  of a black hole is still available.  Especially, the effective metric as a methodology has been applied widely in the non-linear dynamics including moving fluids, Bose-Einstein condensates, superfluids, nonlinear electromagnetism,  and so on (see \cite{effectm} for a review). In this Letter,
  we apply effective metric to study the  black hole entropy arising from massless scalar field with Lorentz violation induced by the coupling to Einstein tensor. Our results show that with Lorentz violation the coupled massless scalar field contributes to the classical Bekenstein-Hawking term in black hole entropy. The corrected classical Bekenstein-Hawking entropy is not one quarter of the event horizon area of the original background black hole, but of a corresponding effective metric. It means that the classical Bekenstein-Hawking entropy depends not only on the black hole parameter, but also on the coupling which gives rise to Lorentz violation.


In the standard coordinate, the metric for a static black hole
spacetime can be expressed as
\begin{eqnarray}
ds^2&=&-fdt^2+\frac{1}{f}dr^2+r^2
d\theta^2+r^2\sin^2{\theta}d\phi^2.\label{m1}
\end{eqnarray}
Here $f$ is a function of the polar coordinate $r$, which can be
written as $f=H(r)(r-r_H)$ for a non-extremal static black hole
spacetime and $H(r)$ is a continuous function of $r$. The quantity $r_H$ is event horizon radius of a black hole. The Einstein tensor $G_{\mu\nu}$ for the metric
(\ref{m1}) has a form
\begin{eqnarray}
G_{\mu\nu}=\left(\begin{array}{cccc}
-a(r)f& & &\\
&\frac{a(r)}{f}&&\\
&&-b(r)r^2&\\
&&&-b(r)r^2\sin^2\theta
\end{array}\right),
\end{eqnarray}
with
\begin{eqnarray}
a(r)=\frac{1-f-f'r}{r^2},\;\;\;\;\;\;\;\;\;\;\;b(r)=\frac{f''r+2f'}{2r},\label{ab0}
\end{eqnarray}
where the prime $'$ denotes the derivative with respect to $r$.

The action of a massless scalar field coupling to Einstein tensor in
the curve spacetime can be described by
\begin{eqnarray}
S&=&\int d^4x\sqrt{-g}\bigg[\frac{R}{16\pi G}+\frac{1}{2}g^{\mu\nu}\partial_{\mu} \psi\partial_{\nu}\psi\nonumber\\&&+\frac{\alpha}{2}G^{\mu\nu}\partial_{\mu} \psi\partial_{\nu}\psi\bigg], \label{act1}
\end{eqnarray}
where the term $\frac{\alpha}{2}G^{\mu\nu}\partial_{\mu} \psi\partial_{\nu}\psi$ represents
the coupling  between scalar field $\psi$ and Einstein tensor $G^{\mu\nu}$. The factor
$\alpha$ is a coupling constant with dimensions of length squared.  Varying
the action (\ref{act1}) with respect to $\psi$ , one can obtain the
corrected wave equation of a scalar field coupling to Einstein tensor
\begin{eqnarray}
\frac{1}{\sqrt{-g}}\partial_{\mu}[\sqrt{-g}( g^{\mu\nu}+\alpha  G^{\mu\nu} )\partial_{\nu} \psi]=0.\label{acts1}
\end{eqnarray}
Separating $\psi=e^{-iEt}\Psi(r)Y_{lm}(\theta,\phi)$,  we obtain
the radial equation for the coupled scalar field
in the static black hole spacetime (\ref{m1})
\begin{eqnarray}
&&\bigg[1+\alpha
a(r)\bigg]f^{-1}E^2\Psi(r)+\frac{1}{r^2}\frac{d}{dr}\bigg[r^2f\bigg(1+\alpha
a(r)\bigg)\nonumber\\&&\frac{d\Psi(r)}{dr}\bigg]-\bigg(1-\alpha
b(r)\bigg)\frac{l(l+1)}{r^2}\Psi(r)=0.\label{wv-sc}
\end{eqnarray}
The radial equation (\ref{wv-sc}) can be reduced to  that in the
usual case of the scalar field without coupling to Einstein tensor
as the coupling constant $\alpha\rightarrow 0$. Here, we consider
only the case with $a(r)\neq-b(r)$ because when $a(r)=-b(r)$ the
coupling does not affect actually the radial equation of the scalar
field propagating in the spacetime under WKB approximation. For the
case with $a(r)\neq-b(r)$, the emergence of the coupling constant
$\alpha$ in the radial equation will change the dynamical behavior
of the scalar field and affect the quantum entropy of the black
hole. However, after some careful analysis, we find that in
this case the coupling term modifies the dispersion relations of the
massless scalar field, which results in that the propagation
velocity of the massless scalar field in the background spacetime
becomes
\begin{eqnarray}
v_{sc}&=&\frac{dl_0}{d\tau}=\sqrt{\frac{g_{rr}p^{r}p^{r}
+g_{\theta\theta}p^{\theta}p^{\theta}
+g_{\phi\phi}p^{\phi}p^{\phi}}{-g_{tt}p^{t}p^{t}}}\nonumber\\&\approx&
\sqrt{1-\frac{\alpha [a(r)+b(r)]l(l+1)}{[1-\alpha
b(r)]r^2E^2}}.\label{vc}
\end{eqnarray}
Here $l_0$ and $\tau$ are the proper distance and time in the
background spacetime, respectively.
$p_{\mu}=\frac{dx^{\mu}}{d\lambda}$ is the momentum of the scalar
field where $\lambda$ is the affine parameter. For the case
$a(r)\neq-b(r)$, one can find that the velocity of a coupled
massless scalar field propagating in the spacetime depends on the
coupling strength between the field and Einstein tensor. The coupled
massless scalar field is superluminal for $\alpha<0$ and is
subluminal for $\alpha>0$. It is well known that the usual event
horizon of black hole is defined originally by the null hypersurface
from which the light can not be escaped, which implies that the
event horizon for the light is not coincide with the event horizons
for the superluminal or subluminal particles. The modification of
the dispersion relations and the appearance of superluminal
phenomenon means that Lorentz invariance is broken for the massless
scalar field coupling to Einstein tensor. It implies that the usual
method for calculating quantum entropy of black hole arising
material field is invalid because the statistical mechanics must be
corrected in the case with the broken Lorentz invariance \cite{LI}.

Although the propagation velocity of the scalar field coupling to Einstein
tensor is not the speed of light in the original static black hole spacetime
(\ref{m1}), we find that it can propagate with ``light" velocity
in another effective spacetime with the metric $\tilde{g}_{\mu\nu}$, which is defined by rewriting the action (\ref{act1}) as
\begin{eqnarray}
S&=&\int d^4x\sqrt{-g}\left[\frac{1}{2}g^{\mu\nu}\partial_{\mu}
\psi\partial_{\nu}\psi+\frac{\alpha}{2}G^{\mu\nu}\partial_{\mu}
\psi\partial_{\nu}\psi\right]\nonumber\\&=&\int d^4x\sqrt{-\tilde{g}}
\left[\frac{1}{2}\tilde{g}^{\mu\nu}\partial_{\mu} \psi\partial_{\nu}\psi\right].\label{act20}
\end{eqnarray}
In other words, the coupled massless scalar field in the original background spacetime $g_{\mu\nu}$ (\ref{m1}) can be regarded as a free scalar field in the effective spacetime with $\tilde{g}_{\mu\nu}$, which means that the Lorentz invariance is guaranteed formally in this effective spacetime and then the usual method for calculating quantum entropy of black hole is still valid in this case. The effective metric $\tilde{g}_{\mu\nu}$  can be expressed as
\begin{eqnarray}
&&ds^2=-[1-\alpha b(r)]fdt^2+[1-\alpha
b(r)]\frac{1}{f}dr^2\nonumber\\&&+[1+\alpha a(r)]r^2 d\theta^2+[1+\alpha
a(r)]r^2\sin^2{\theta}d\phi^2.\label{m2}
\end{eqnarray}
Obviously, the effective metric is quite different from the original one.
Due to its dependence on the coupling constant $\alpha$, the effective metric coefficients are different for the coupled scalar fields with different coupling strength. From the definition of the null hypersurface $\tilde{g}^{\mu\nu}\frac{\partial \tilde{f}}{\partial x^{\mu}}\frac{\partial \tilde{f}}{\partial x^{\nu}}=0$,
it is easy to find that the event horizon radius for the scalar field
 satisfies $\tilde{g}^{rr}=\frac{f}{1-\alpha
b(r)}=0$ and is still $r=r_H$ in the static effective spacetime
(\ref{m2}). However, the area of the event horizon is modified in
this case, which could change the properties of the black hole
entropy. With this effective metric, it is convenient for us to make
a comparison among quantum entropy for the scalar field with
different coupling parameter. From the quantity
$R_{\mu\nu\rho\sigma}R^{\mu\nu\rho\sigma}=\frac{\mathcal{R}(r)}{[1+\alpha
a(r)]^4[1-\alpha b(r)]^6r^4}$ (where $\mathcal{R}(r)$ is a function
of $r$), we can find that there exist two extra singularities which
is caused entirely by the coupling between the scalar field and
Einstein tensor in the original background spacetime. Considering
that the scalar field should propagate continuously in the region
outside the event horizon, we make a constraint condition that two
singularities lie inside the event horizon $r_H$ so that they do not
affect the propagation of the scalar field in the spacetime
$\tilde{g}_{\mu\nu}$.

From Eq.(\ref{act20}), one can find that the equation of motion of a coupled mass scalar field can be rewritten as the form of a free Klein-Gordon equation in the effective spacetime
\begin{eqnarray}
\frac{1}{\sqrt{-\tilde{g}}}\partial_{\mu}[\sqrt{-\tilde{g}}
( \tilde{g}^{\mu\nu})\partial_{\nu} \psi]=0.\label{me2}
\end{eqnarray}
Separating the variable
$\psi=e^{-iEt}\Psi(r)Y_{lm}(\theta,\phi)$ and inserting the effective metric (\ref{m2}) into  equation (\ref{me2}), we find that the radial equation has the same form with the previous radial equation (\ref{wv-sc}). However, the Lorentz invariance is guaranteed in this case and we can compute the quantum entropy of black hole by usual 't Hooft ``brick-wall" method \cite{Hooft}. From the radial equation
(\ref{wv-sc}), one can obtain the $r$-dependent radial wavenumber
\begin{eqnarray}
k^2(r,l,E)&\equiv & p^2_r= f^{-2}\bigg[E^2-f\bigg(\frac{p^2_{\theta}}{g_{\theta \theta}}
+\frac{p^2_{\phi}}{g_{\phi \phi}}\bigg)\bigg]\nonumber\\
&=& f^{-2}\bigg[E^2-f\frac{l(l+1)}{r^2}
\bigg(\frac{1-\alpha b(r)}{1+\alpha a(r)}\bigg)\bigg].\nonumber\\ \label{wns}
\end{eqnarray}
From the ``brick-wall" model, one can find that the boundary conditions
of the scalar field $\Psi(r)$ are
\begin{eqnarray}
\Psi(r_H+h)=0,\;\;\;\;\;\;\;\;\Psi(L)=0,
\end{eqnarray}
where $h\ll r_H$ and $L\gg r_{H}$. Here $h$ and $L$ are the brick-wall
cut-off parameters introduced to eliminate the ultraviolet and
infrared divergences, respectively.  Adopting to the WKB
approximation, we have $\Psi(r)\approx e^{iS(r)}$. Here, the
solution $\Psi(r)$ is regarded to be stationary with regard respect to the
radial variable throughout the spatial manifold and the amplitude is
assumed to be a slowly varying function of the radial coordinate
$r$.

According to the semi-classical quantization condition, the total number of radial modes is given by
\begin{eqnarray}
n_r(r,l,E)=\frac{1}{\pi}\int^{L}_{r_H+h} k(r,l,E)dr.\label{n0}
\end{eqnarray}
The total number of the modes with energy $E$ can be obtained by integrating over the volume of phase space \cite{RBM}
\begin{eqnarray}
&&n(E)=\frac{1}{(2\pi)^2} \int d\theta d\phi \int n_r(r,l,E) dp_{\theta}dp_{\phi}\nonumber\\
&&=\frac{1}{4\pi^3}\int^{L}_{r_H+h} dr \int d\theta d\phi \int k(r,l,E) dp_{\theta}dp_{\phi}.\nonumber\\
&&=\frac{1}{\pi}\int^{L}_{r_H+h}
f^{-1}dr\times \nonumber\\
&& \int
\sqrt{E^2-f\bigg[\frac{l(l+1)}{r^2} \bigg(\frac{1-\alpha
b(r)}{1+\alpha a(r)}\bigg)\bigg]}(2l+1)dl.\nonumber\\ \label{n}
\end{eqnarray}
For an equilibrium ensemble
of states of the coupled field,
the free energy is given by
\begin{eqnarray}
\beta F&&=\int \ln(1-e^{-\beta E})dn(E)\nonumber\\&&=-\beta\int \frac{n(E)}{e^{\beta E}-1}dE.\label{FE}
\end{eqnarray}
Substituting  the wavenumber (\ref{wns}) and the number of the modes (\ref{n}) into equation (\ref{FE}), we can obtain the free energy for the coupled scalar field
\begin{eqnarray}
 F&=&-\frac{1}{\pi}\int^{L}_{r_H+h}
f^{-1}dr\int^{\infty}_{0} \frac{1}{e^{\beta E}-1}dE
\nonumber\\&\times&\int
\sqrt{E^2-f\bigg[\frac{l(l+1)}{r^2} \bigg(\frac{1-\alpha
b(r)}{1+\alpha a(r)}\bigg)\bigg]}(2l+1)dl\nonumber\\
&=&-\frac{2\pi^3}{45\beta^4}\int^{L}_{r_H+h}
 \frac{r^2}{f^2}\bigg(\frac{1+\alpha a(r)}{1-\alpha b(r)}\bigg)dr.\label{FEs1}
\end{eqnarray}
Letting $r=r_H + x$ and ignoring the effect from the boundary $L'$, one can find that the free energy obeys
\begin{eqnarray}
&F&\approx
-\frac{2\pi^3}{45\beta}\int^{L'}_{h}\frac{(r_H+x)^2[1+\alpha
a(r_H+x)]}{x^2H(r_H+x)^2[1-\alpha b(r_H+x)]} dx
\nonumber\\&=&-\frac{2\pi^3r^2_H[1+\alpha
a(r_H)]}{45\beta^4H(r_H)^2[1-\alpha b(r_H)]h}+\frac{2\pi^3r_H
}{45\beta^4H(r_H)^2}\nonumber\\&\times&\bigg\{\frac{2[1+\alpha
a(r_H)][H(r_H)-r_HH'(r_H)]}{[1-\alpha
b(r_H)]H(r_H)}\nonumber\\&&+\alpha
r_H\bigg[\frac{a(r_H)+b(r_H)}{1-\alpha b(r_H)}\bigg]'\bigg\}\ln h.
\label{FEs3}
\end{eqnarray}
Then the statistical-mechanical entropy of the black hole arising from the scalar field coupling to Einstein tensor is given by
\begin{eqnarray}
&&S=\beta^2\frac{\partial
F}{\partial\beta}\bigg|_{\beta=\beta_H}=\frac{r^2_HH(r_H)}{360h}
\bigg[\frac{1+\alpha
a(r_H)}{1-\alpha
b(r_H)}\bigg]\nonumber\\&&-\frac{r_HH(r_H)}{360}\bigg\{\frac{2[H(r_H)-r_HH'(r_H)]}{H(r_H)}\times\nonumber\\&&\frac{[1+\alpha
a(r_H)]}{[1-\alpha
b(r_H)]}+\alpha r_H\bigg[\frac{a(r_H)+b(r_H)}{1-\alpha b(r_H)}\bigg]'\bigg\}\ln h.\nonumber\\&&\label{Ssco}
\end{eqnarray}
For the effective spacetime (\ref{m2}), the surface gravity at the horizon $r_H$ is $\kappa=-\frac{1}{2}\sqrt{\frac{-1}{\tilde{g}_{tt}\tilde{g}_{rr}}}
\frac{d\tilde{g}_{tt}}{dr}\bigg|_{r\rightarrow r_H}$  and then $\beta_H=\frac{2\pi}{\kappa}=\frac{4\pi}{H(r_H)}$. The proper distance $\delta$ from the horizon
$r_H$ to the $r_H+h$ becomes
\begin{eqnarray}
\delta=\int^{r_H+h}_{r_H}\sqrt{\tilde{g}_{rr}}dr=2\sqrt{\frac{[1-\alpha b(r_H)] h}{H(r_H)}}.
\end{eqnarray}
Defining $\delta^2=\frac{2\epsilon^2}{15}$, we find the entropy (\ref{Ssco}) can be rewritten as
\begin{eqnarray}
&&S=\frac{\tilde{A}_H}{48\pi\epsilon^2}-\frac{r_HH(r_H)}{360}
\bigg\{\frac{2[H(r_H)-r_HH'(r_H)]}{H(r_H)}\times\nonumber\\&&\frac{[1+\alpha
a(r_H)]}{[1-\alpha
b(r_H)]}+\alpha r_H\bigg[\frac{a(r_H)+b(r_H)}{1-\alpha b(r_H)}\bigg]'\bigg\}\ln h. \label{Ssc1}
\end{eqnarray}
Here $\tilde{A}_H=\pi r^2_H[1+\alpha a(r_H)]$ is the event horizon area of the
effective metric (\ref{m2}) rather than the original one (\ref{m1}),  which depends on
the coupling between the scalar field and the Einstein tensor. Obviously, the first term in
the black hole entropy (\ref{Ssc1}) can be expressed as one quarter of the
 event horizon area of the effective metric by the renormalization of the Newton gravity
 constant $G$. This means that the coupling constant changes not only the logarithmic
 correction term, but also  the classical Bekenstein-Hawking term in black hole
 entropy, which is quite different from that in the case without the coupling
 between the scalar field and the Einstein tensor. With increase of the coupling
 strength, the event horizon area $\tilde{A}_H$ and the black hole entropy
 increase for the subluminal field with the positive coupling constant and decrease
 for the superluminal field with the negative one. It can be understood by a fact
 that the superluminal signal can give us more information from a black hole.
Although the form of the corrected Bekenstein-Hawking entropy is
very similar to those of some black holes carried certain a ``hair",
we must point out that the effective metric is \emph{not} a solution of the
field equation and the original background black hole \emph{does not} carry
any new ``hair". Thus, our study do not violate the ``no hair"
theorem. The non-geometrical terms in the entropy formula originates
from the Lorentz violation induced by the coupling to Einstein
tensor. The main reason is that Lorentz violation leads to that in
the original background spacetime the maximum propagation velocity
of the massless scalar field  is not the speed of light and then the
usual event horizon for light is not again the horizon for the
massless scalar field coupling to the Einstein tensor.

Summary, we have studied quantum entropy of a static black
hole arising from the massless scalar field with Lorentz violation
induced by the coupling to Einstein tensor. Our results show that
there exist  contributions  from the coupled massless scalar field
to the classical Bekenstein-Hawking term in  black hole entropy.
The corrected classical Bekenstein-Hawking entropy is not one
quarter of the event horizon area of the original background black
hole, but of a corresponding effective metric related to the
coupling. It means that the classical Bekenstein-Hawking entropy
depends not only on the black hole parameter, but also on the
coupling giving rise to Lorentz violation.

This work was  partially supported by the National Natural Science
Foundation of China under Grant No.11275065,  No. 11475061,
the construct program of the National Key Discipline, and the Open Project Program
of State Key Laboratory of Theoretical Physics, Institute of Theoretical Physics,
Chinese Academy of Sciences, China (No.Y5KF161CJ1).


\vspace*{0.2cm}
\begin{thebibliography}{99}

\bibitem{Beken1} J. D. Bekenstein, Nuovo Cimento Lett. {\bf4}, 737  (1972);
J. D. Bekenstein,  Phys. Rev. D. {\bf 7}, 2333 (1973).
\bibitem{Hawk1}  S. W. Hawking,  Nature {\bf248}, 30 (1974);
S. W. Hawking,    Commun. Math. Phys. {\bf43}, 199 (1975).
\bibitem{Beken2}  J. D. Bekenstein, Phys. Rev. D {\bf9}, 3292 (1974).

\bibitem{Hooft}  G. 't Hooft, Nucl. Phys. B {\bf256}, 727 (1985).

\bibitem{ae1} A. G. Riess et al., Astron. J. {\bf116}, 1009 (1998) ; A. G.
Riess et al., Astrophys. J. {\bf659}, 98 (2007).

\bibitem{ae2} P. de Bernardis et al., Nature (London) {\bf404}, 955 (2000)
\bibitem{ae3} S. Perlmutter et al., Astrophys. J. {\bf517}, 565 (1999); Astrophys. J. {\bf598},102 (2003).
\bibitem{ae4} D. J. Eisenstein et al., Astrophys. J. {\bf633}, 560 (2005);
D. N. Spergel et al., Astrophys. J. Suppl. Ser.{\bf170}, 377
(2007).

\bibitem{Sushkov} S. V. Sushkov, Phys. Rev. D {\bf80}, 103505 (2009).
\bibitem{Gao} C. J. Gao, J. Cosmol. Astropart. Phys. {\bf06}, 023 (2010).
\bibitem{Sarida} E. N. Saridakis and S.V. Sushkov, Phys. Rev. D {\bf81},
083510 (2010).
\bibitem{CG1}C. Germani and A. Kehagias, Phys. Rev. Lett. {\bf105}, 011302 (2010).

\bibitem{James} J. B. Dent, S. Dutta, E. N. Saridakis and J. Q, Xia, J. Cosmol. Astropart. Phys.{\bf11}, 058 (2013).

\bibitem{sb2010} S. Chen and J. Jing, Phys. Lett. B {\bf691}, 254 (2010); Phys. Rev. D {\bf82},084006 (2010).


\bibitem{Mas} M. Minamitsuji, Phys. Rev. D {\bf89}, 064025 (2014);  Phys. Rev. D {\bf89}, 064017 (2014).
\bibitem{BHC2} A. Anabalon, A. Cisterna and J. Oliva, Phys. Rev. D {\bf89}, 084050 (2014).
\bibitem{Theod} T. Kolyvaris, G. Koutsoumbas, E. Papantonopoulos and G. Siopsis,  J. High Energy Phys.{\bf11},133 (2013).
    
\bibitem{BHC1} C. Germani, L. Martucci and P. Moyassari, Phys. Rev. D {\bf85}, 103501 (2012).


    
\bibitem {LI0} David Mattingly, Living Rev. Rel. {\bf 8}, 5 (2005).
\bibitem {LI} D. Colladay and P. McDonald, Phys. Rev. D {\bf70},  125007 (2004).
\bibitem {effectm} C. Barcelo, S. Liberati, and M. Visser, Living Rev. Rel.{\bf8},12 (2005); arXiv: gr-qc/0505065.

\bibitem{RBM} R. B. Mannt, L. Brasovt and A. Zelnikov, Class. Quantum Grav. {\bf9}, 1487 (1992).

\end{thebibliography}
\end{document}